\documentclass[preprint]{ptephy_v1}

\preprintnumber{XXXX-XXXX} 
\usepackage{hyperref}

\usepackage{amsmath} 
\usepackage{amsthm} 
\usepackage{hyperref} 
\usepackage{graphics} 
\usepackage{algorithmic} 
\usepackage{url} 
\usepackage{caption}
\usepackage{subcaption}
\usepackage[table]{xcolor}
\usepackage[section]{placeins}
\usepackage{lineno}

\begin{document}

\title{Metrology requirements for the integrated luminosity measurement using small-angle Bhabha scattering at ILC}


\author[1]{Ivan Smiljanić}
\affil{Vinca Institute of Nuclear Sciences - National Institute of the Republic of
Serbia, University of Belgrade, M. Petrovica Alasa 12-14, Belgrade, Serbia \email{i.smiljanic@vin.bg.ac.rs}}

\author[1]{Ivanka Božović}

\author[1]{Goran Kačarević}

\author[2]{Mirko Radulović} 
\affil{University of Kragujevac, Faculty of Science, 34000 Kragujevac, Serbia}

\author[2]{Jasna Stevanović} 


\begin{abstract}%
Precision measurement of the integrated luminosity $\mathcal{L}$ at future Higgs factories, including ILC, is of crucial importance for the cross-section measurements, and in particular for the line-shape measurements at the Z-pole. Since there is no up-to-date estimate of the integrated luminosity uncertainties arising from metrology effects at ILC, here we review the metrology requirements for the targeted precision of $\mathcal{L}$ at foreseen ILC center-of-mass energies: 91.2 GeV, 250 GeV, 500 GeV and 1 TeV, using small-angle Bhabha scattering.
\end{abstract}


\maketitle

\section{Introduction}
\label{sec:intro}
    The realistic target for integrated luminosity ($\mathcal{L}$) precision at future high energy electron-positron colliders is set by the requirements of their physics programs, primarily aiming to study the properties of a Higgs boson in the operation mode at around 250 GeV center-of-mass energy. For the most of these measurements, relative systematic uncertainty of the integrated luminosity of $\sim 10^{-3}$ should suffice. The same condition of uncertainty of the integrated luminosity can be assumed for $e^+ e^-$ collider with higher energies up to TeV, which is accessible as an upgrade of the International Linear Collider (ILC) \cite{ILCTDR3}. At lower center-of-mass energies dedicated for precision EW studies, the targeted luminosity precision is set to $\sim 10^{-4}$, primarily from the line-shape measurements at the Z resonance and the W boson mass measurement from the W-pair production line shape at the threshold \cite{1}. 
    
The feasibility of such required precision is challenged by numerous requirements related to manufacturing, positioning and performance of the luminosity monitor, in addition to the uncertainties of beam properties (transverse and longitudinal sizes, energy and beam delivery to the interaction point), collectively referred to as metrology. These effects impact the available phase space for detection of the final state particles and consequently their count, introducing systematic uncertainty in the integrated luminosity measurement.

Since the monitoring systems for most of these effects are under development at ILC, this paper aims to provide tolerable margins for the effects under study from the point of view of the integrated luminosity precision goals. Each effect is set to contribute to the relative uncertainty of the integrated luminosity as $1 \cdot 10^{-3}$ ($1 \cdot 10^{-4}$) at the center-of-mass energy of 250 GeV and above (at 91.2 GeV), limiting the overall contribution from metrology to the relative uncertainty of $\mathcal{L}$ to less than $3.3 \cdot 10^{-3}$ ($3.3 \cdot 10^{-4}$), since many of the precision margins presented here can be easily surpassed with the existing technologies. As pointed out in the Focus topics document for the ECFA study on Higgs/top/EW factories \cite{3}, the impact of metrology on the integrated luminosity measurement at ILC has not been estimated yet, up to the generic study for a long time ago proposed TESLA collider \cite{4} discussed in \cite{5}.

This study is based on the small angle Bhabha scattering (SABS), the conventionally used process as a large rate, almost pure QED process with the theoretical uncertainty of $3.7 \cdot 10^{-4}$ recently revised for the LEP analyses \cite{2}, although neither NLO electroweak corrections nor the multi-fermion production are implemented in the existing Bhabha MC tools \cite{3}. Once the SABS cross-section ($\sigma$) is known in a certain phase space, the number of Bhabha counts (N) can be used to measure the integrated luminosity as $\mathcal{L} = \frac{N}{\sigma}$. The uncertainty of Bhabha's count is translated as the uncertainty of $\mathcal{L}$.

The structure of the paper is the following: Section \ref{sec:sec2} brings a brief overview of instrumentation of the very forward region at ILC and the considered systematic effects, while Section \ref{sec:sec3} brings limits on the detector and beam-related parameters at center-of-mass energies foreseen at ILC, assuming the maximal contribution per effect of $10^{-3}$ ($10^{-4}$) at the center-of-mass energy of 250 GeV and above (at 91.2 GeV).

\section{Luminosity measurement at ILC}
\label{sec:sec2}
There was an extensive work by the FCAL Collaboration on luminometer design, performance and prototyping, documented in \cite{6}, \cite{7} and \cite{8}. The FCAL Collaboration has proposed instrumentation of the very forward region of the International Large Detector (ILD)  at ILC \cite{9} with two silicon-tungsten sandwich calorimeters, one for the fast luminosity estimate and measurement of the beam properties (Beam Cal) and the other for measurement of the integrated luminosity (LumiCal). The feasibility of realization of these devices has been proven in several test beam campaigns, demonstrating performance of a calorimeter prototype \cite{6}. Systematic uncertainties from LumiCal performance are discussed in \cite{7}, while the impact of beam related effects on reconstruction of the luminosity spectrum is discussed in \cite{8}. As pointed out, systematic effects from metrology have not been quantified yet.     

\subsection{Instrumentation of the very forward region}
\label{sec:sec2.1}
\begin{figure}[!h]
\centering\includegraphics[width=.50\textwidth]{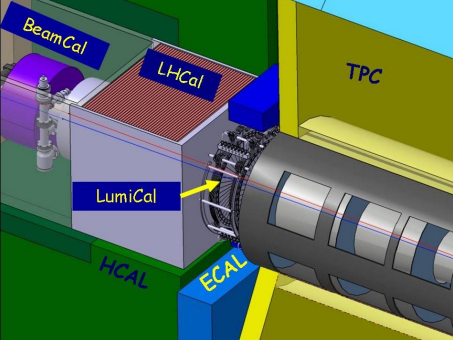}
\caption{Layout of the very forward region of a detector at ILC, with LumiCal placed $\mathrm{2500 \: mm}$ from the IP.}
\label{fig_1}
\end{figure}

The layout of the very forward region at ILD is given in Figure \ref{fig_1} \cite{7}. At 2500 mm from the interaction point (IP), LumiCal will be  centered around the outgoing beam at 7 mrad polar angle with respect to the z-axis, preserving the symmetry of head-on collisions at ILC with the crossing angle of 14 mrad. The geometrical aperture of the LumiCal is 31-77 mrad, while the sampling term is constant in the range of  41-67 mrad, which defines the fiducial volume (FV) \cite{7}. LumiCal consists of 30 subsequent silicon-tungsten layers providing longitudinal coverage for showers developed from high-energy Bhabhas. Each silicon layer (including the silicon sensor, capton HV and fan-out, together with the supportive structure) is 640 $\mathrm{\mu m}$ thick and the thickness of the tungsten absorber is $3.5$ $mm$ each, which corresponds to one radiation length. To provide precision position measurement in the transverse plane, silicon sensors are finely segmented into pads (48 azimuthal sectors and 64 radial rings), each with 1.8 mm pitch.   

The study presented in this paper relies only on the size of the fiducial volume and its effective change due to the effects to be described (Section 2.3) without taking into consideration detector performance based on the given choice of technology. 

\subsection{Event generation}
\label{sec:sec2.2}
We have simulated 10 million small angle Bhabha scattering events using the BHLUMI V4.04 Bhabha event generator \cite{11} for each energy considered. Bhabha events are generated from 20 mrad to 200 mrad, to allow events with non-collinear final state radiation to contribute. Initial state radiation is also considered while beam-beam interactions are neglected. Also, we do not consider electromagnetic deflection of the final state due to interactions with the field of the corresponding outgoing beam. These effects have been discussed in \cite{8}. Also, in \cite{8} it has been shown that LumiCal at ILC will not be significantly affected by the four-fermion background from the Landau-Lifshitz type of processes (‘two-photon’ exchange), so we do not consider it here either.

\begin{figure}[!h]
\centering\includegraphics[width=.50\textwidth]{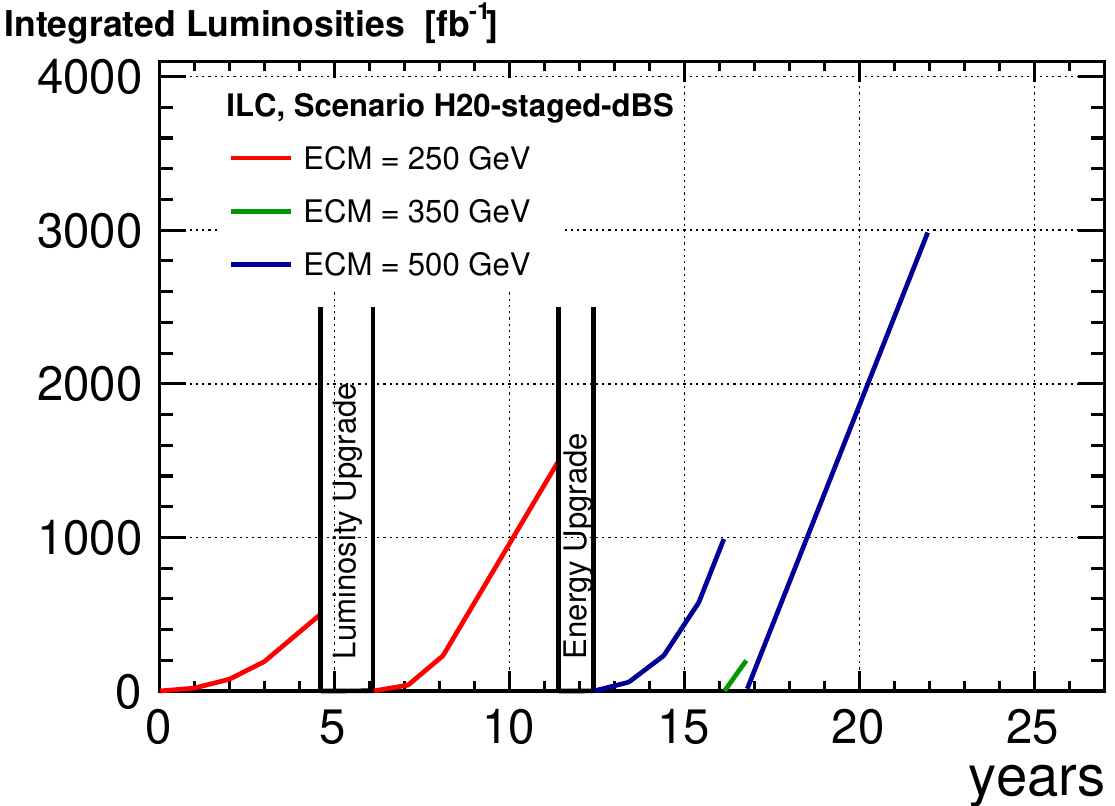}
\caption{Integrated luminosity vs. ILC running time in the H-20 operating scenario.}
\label{fig_2}
\end{figure}

Bhabha cross-sections in the considered range of polar angles are: 161.5 nb, 22.4 nb, 4.5 nb and $\mathrm{0.79 \: nb}$, at $\mathrm{91.2 \: GeV}$ (Z-pole), $\mathrm{250 \: GeV}$, $\mathrm{500 \: GeV}$ and $\mathrm{1 \: TeV}$ center-of-mass energy, respectively.  Expected integrated luminosities assume the so called H-20 staged run  \cite{12} of ILC  at 250 GeV and 500 GeV with the corresponding integrated luminosities of 2 ab$^{-1}$ and 4 ab$^{-1}$, to be complemented with 8 ab$^{-1}$ of data at 1 TeV center-of-mass energy and $\mathrm{100 \: fb^{-1}}$ at the 91.2 GeV. Figure \ref{fig_2} \cite{Fujii:2017vwa} illustrates estimated ILC running time and integrated luminosities in the H-20 operating scenario.

We have applied asymmetrical event selection in polar angles (AS1), in a way that at one side of LumiCal a full fiducial volume is considered, while at the other side the inner and outer radii of the fiducial volume are narrowed by 1 mm.  Modification of the counting volume is applied subsequently on the left and right arm of LumiCal, on event-by-event basis. This approach (with some variations between experiments) has been applied at LEP \cite{13}, reducing the net systematic effect arising from left-right asymmetries. This, however, requires that LumiCal arms are positioned on the outgoing beams at colliders with a non-zero crossing angle. It should be noted that the statistical fluctuation of generated samples is of order of $3 \cdot 10^{-4}$ and for this reason we rather illustrate the dependences of the effects under study then attempting to model an effect with a fit.

\subsection{Effects from metrology}
\label{sec:sec2.3}
We are considering the following systematic effects that will be present due to the uncertainties of the LumiCal radial dimensions, positioning with respect to the IP and relative positioning of detector halves, as well as due to LumiCal vibrations in the transverse and longitudinal direction:
\begin{enumerate}
\item uncertainty of the inner radius of the LumiCal's counting volume ($\Delta r_{in}$),
\item uncertainty of the outer radius of the LumiCal's counting volume ($\Delta r_{out}$),
\item RMS of the Gaussian spread of the measured radial shower position with respect
to the true hit position\footnote{True Bhabha position can be measured by placing a Si tracker plane in front of LumiCal with the precision of several microns. In addition, this would provide electron-photon separation.} ($\sigma_{r}$), 
\item uncertainty of the longitudinal distance between left and right halves of LumiCal ($\Delta l$) assuming that both halves are moving simultaneously towards the IP or away from it for $\Delta l/2$,
\item  RMS of the Gaussian distribution of mechanical fluctuations of LumiCal
with respect to the IP in the radial direction ($\sigma_{x_{IP}}$),  
\item  RMS of the Gaussian distribution of mechanical fluctuations of LumiCal
with respect to the IP in the axial direction ($\sigma_{z_{IP}}$), 
\item tilt of LumiCal (both arms) equivalent to a rotation around y-axis for a certain angle (tilt).

\end{enumerate}
Also, the beam bunches are finite in size and interaction may occur anywhere inside the bunches. In addition, beam synchronization may cause longitudinal (axial) displacements of the IP. Thus we consider:

\begin{enumerate}
\setcounter{enumi}{7}

\item radial ($\Delta x_{IP}$) displacements of the interaction point with respect to LumiCal, 
\item axial ($\Delta z_{IP}$) displacements of the interaction point with respect to LumiCal. From $\Delta z_{IP}$, maximal time shift in beam synchronization ($\Delta \tau$) can be determined.

\end{enumerate}

Any asymmetry of the beam energies occurring either on an event-by-event basis as a random variation due to the beam energy spread (BES) or as a permanent bias of one beam energy with respect to the other, will cause a boost (assumed longitudinal\footnote{For the boost to be longitudinal, we assume that the initial state radiation is emitted along the z-axis (in the head-on collisions geometry - LumiCal placed at the outgoing beams) and that the final state radiation is emitted in the direction of a final state electron (positron). }) of the initial and consequently of the final electron-positron system. The boost ($\beta_{z}$) of Bhabha particles will lead to an effective change of the LumiCal's acceptance seen from the final state electron (positron) reference frame and to the consequent change of the Bhabha count. We thus consider two additional systematic effects:

\begin{enumerate}
\setcounter{enumi}{9}

\item RMS of the Gaussian distribution of the beam energy spread ($\sigma_{E_{BS}}$) and
\item difference in energy ($\Delta E$) between the colliding beams.

\end{enumerate}

Every effect is individually considered to lead to the overall relative change of count of $10^{-4}$ ($10^{-3}$) at 91.2 GeV (higher center-of-mass energies). In this respect, the maximal tolerance is derived for the parameters under study.

\section{Precision requirements for the integrated luminosity measurement}
\label{sec:sec3}
Precision requirements for the integrated luminosity measurement translates to a similar precision requirements on the center-of-mass energy, since the Bhabha cross-section scales  with the center-of-mass energy as $\propto 1/s$ . In the range of the ILC energies, the center-of-mass energy should be known at the level of several MeV at 91.2 GeV, to a few hundred MeV at higher center-of-mass energies. The latter seems to be feasible with experimental reconstruction of di-muon processes at 250 GeV center-of-mass energy \cite{14}, providing $\sqrt{s}$  absolute precision of the order of the beam energy spread of 0.19\% \cite{Adolphsen:2013jya}. At 91.2 GeV, the situation seems to be more challenging. However, for the cross-section measurements of the s-channel processes with the similar cross-section dependence on $s$, uncertainties arising from the limited knowledge of the center-of-mass energy will cancel out in the luminosity estimate. 

A whole set of effects will stem from the detector capability to measure energy and position of scattered Bhabha electrons (positrons). We do not discuss these effects further, since the performance of LumiCal in terms of energy and position reconstruction is not yet fully known. In \cite{6} it is demonstrated that the prototype of LumiCal with six detector planes can reconstruct the signal hits with the resolution of $\mathrm{440 \: \mu}m$ in the front plane. In  this section we shall consider the required resolution at various ILC center-of-mass energies.

\subsection{Z-pole}
\label{sec:sec3.1}

Figures \ref{fig3} (a) to \ref{fig13} (a) illustrate the dependence of the relative change of Bhabha count ($\Delta N/N$) on the metrology effects listed as 1 to 11 respectively. As the maximal value for each effect we take the one corresponding to $\Delta\mathcal{L}/\mathcal{L} = \Delta N/N = 10^{-4}$. If there is more than one value of the metrological uncertainty corresponding to this luminosity (count) precision, the smaller tolerance is taken (i.e. $\Delta r_{in}, \Delta l, \Delta x_{IP}, \Delta z_{IP}$). All the fluctuations that can be seen on Figures \ref{fig3} to \ref{fig13} for all energies are of statistical nature, caused by the limited size of samples. 

As can be seen from Figure \ref{fig3} (a), illustrating the contribution of inner radius uncertainty to uncertainty of the integrated luminosity, $\Delta r_{in}$, the most challenging requirement at the Z-pole comes from the precision of the inner radius of the counting volume of a luminometer. If events are counted in a symmetrical way considering the full detector fiducial volumes on both sides (indicated as 'Fiducial' on Figures \ref{fig3} and \ref{fig4} (a)), a precision of the inner aperture of the counting volume of $\sim$ 3 $\mu $m is required. The asymmetrical counting (described in Section 2.2) compensates for the variations in counting caused by uncertainty of the inner dimension of the counting volume, by introducing much larger ($\sim$ 1 mm) variations of the inner and outer radius, thus relaxing the maximal uncertainty of the inner radius of the counting volume to $\sim$ 20 $\mu$m (Figure \ref{fig3} (a)) (indicated as AS1 on Figures \ref{fig3} to \ref{fig13}). A full simulation of LumiCal response to showers produced by SABS is needed to understand how the uncertainty of the inner aperture of the device transfers to the fiducial volume as the counting region. 

Other metrology effects related to LumiCal positioning that can be seen from Figures \ref{fig4} (a) to \ref{fig13} (a)  are ranging from a few hundreds of micrometers to several millimeters; these should not be an issue for contemporary laser positioning systems based on frequency scanning interferometry. In a system implemented in the ATLAS experiment, 1 $\mu$m absolute precision in positioning can be reached over 1 m distance \cite{15}. As can be seen from Figure \ref{fig7},  dissipation of hits' true position in the LumiCal front plane should not be larger than $\sim$ 300 $\mu $m. This can be mitigated either by finer silicon sensor segmentation, or, more easily, by placing a silicon tracking plane in front of the LumiCal that would in addition provide electron-photon separation. LumiCal should be aligned with the outgoing beam with a tilt no larger than $\sim 14$ mrad (Figure \ref{fig9} (a)). A dedicated luminometer positioning system would have to be developed to ensure the feasibility of this precision. If the axial displacement of the IP was caused by beam synchronization (Figure \ref{fig11} (a)), the time shift between beams should not be larger than $\sim$ 13 ps . Beam energy spread at 91.2 GeV center-of-mass energy should not be larger than $\sim$ 110 MeV (Figure \ref{fig12} (a)), what can be accomplished assuming that the BES is not larger than 0.25\%. Eventual difference in energy of one beam with respect to the other should not exceed $\sim$ 5 MeV, the value of the permanent bias which is highly unlikely to occur. In fact, using the precise J/$\psi$ mass reconstruction targeting 2 ppm at ILC  operating at 91.2 GeV center-of-mass energy, beam energy calibration (including B-field, tracker alignment, material effects, etc.) can be controlled to 10 ppm (better than 500 keV) \cite{snowmass}.

\subsection{Higher center-of-mass energies}
\label{sec:sec3.2}
\begin{figure}[!h]
\centering 
\begin{subfigure}[b]{0.47\textwidth}
\centering
\includegraphics[width=.99\textwidth]{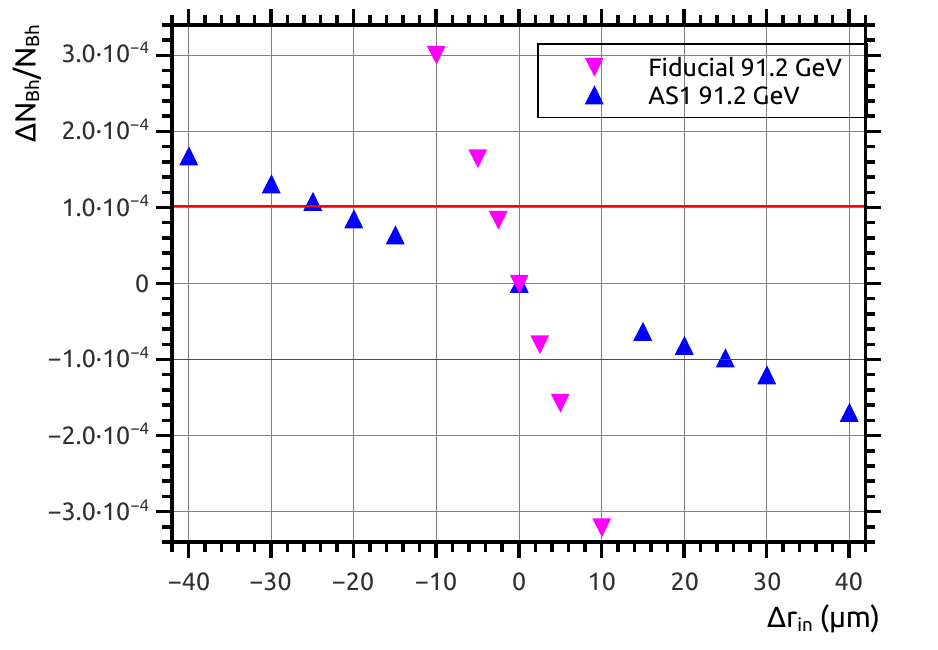}
\caption{}
\end{subfigure}
     \hfill
     \begin{subfigure}[b]{0.47\textwidth}
\centering
\includegraphics[width=.99\textwidth]{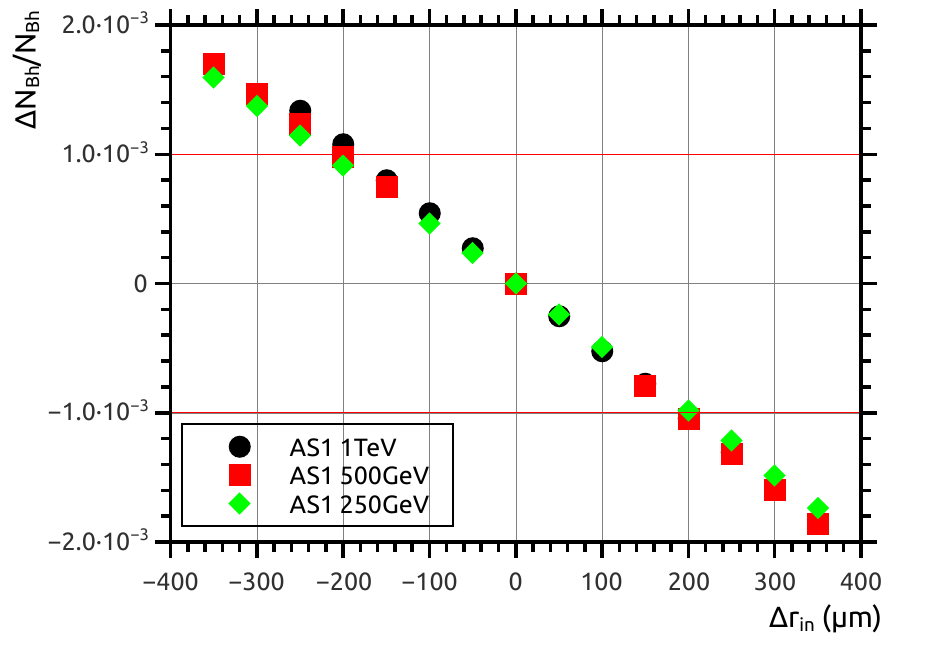}
\caption{}
\end{subfigure}
\caption{\label{fig3} (a) Impact of uncertainty $\Delta r_{in}$ of LumiCal FV inner radius on Bhabha counting, at 91.2 GeV; (b) The same at 250 GeV, 500 GeV and 1 TeV center-of-mass energies for AS1 event selection (Section \ref{sec:sec2.2}).}
\end{figure}

\begin{figure}[!h]
\centering 
\begin{subfigure}[b]{0.47\textwidth}
\centering
\includegraphics[width=.99\textwidth]{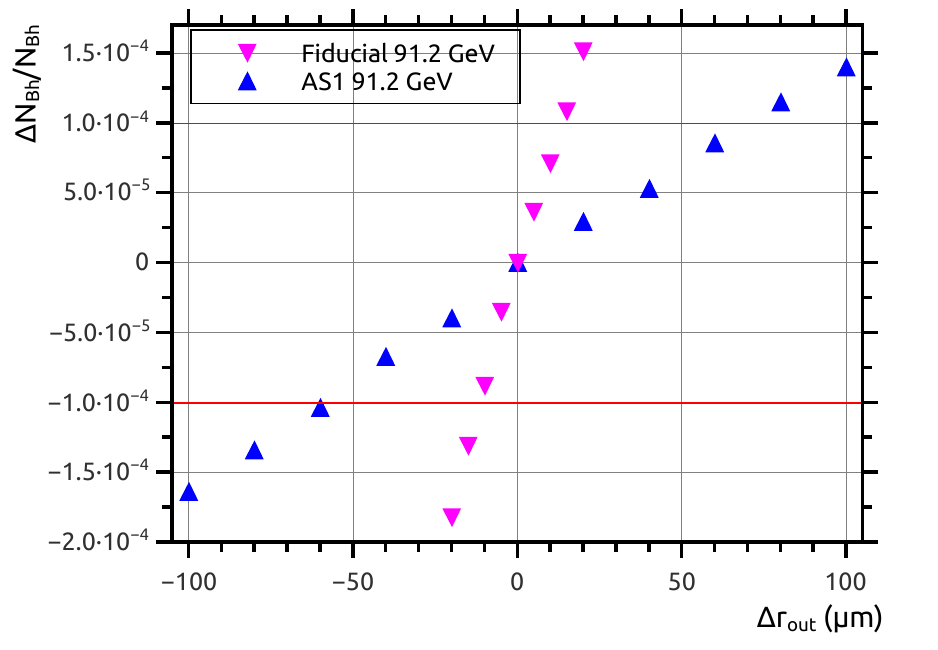}
\caption{}
\end{subfigure}
     \hfill
     \begin{subfigure}[b]{0.47\textwidth}
\centering
\includegraphics[width=.99\textwidth]{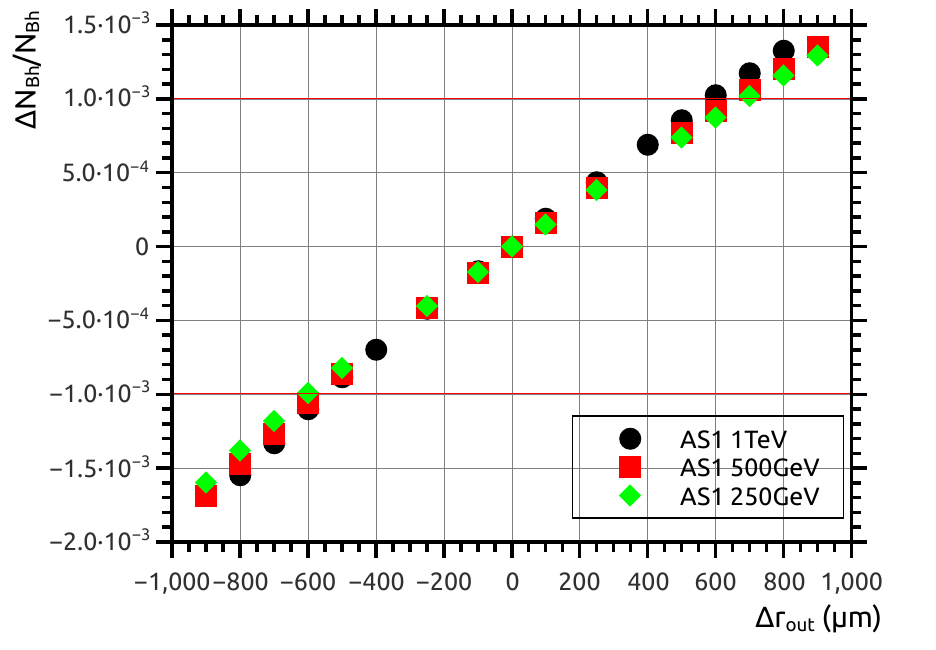}
\caption{}
\end{subfigure}
\caption{\label{fig4} (a) Impact of uncertainty $\Delta r_{out}$ of LumiCal FV outer radius on Bhabha counting, at 91.2 GeV; (b) The same at 250 GeV, 500 GeV and 1 TeV center-of-mass energies for AS1 event selection (Section \ref{sec:sec2.2}).}
\end{figure}

\begin{figure}[!h]
\centering 
\begin{subfigure}[b]{0.47\textwidth}
\centering
\includegraphics[width=.99\textwidth]{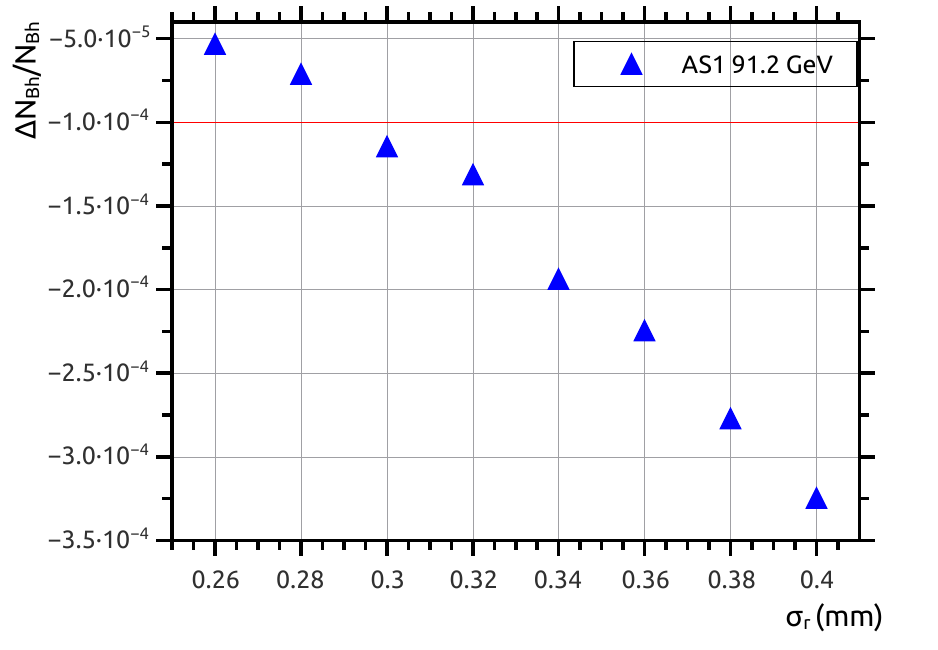}
\caption{}
\end{subfigure}
     \hfill
     \begin{subfigure}[b]{0.47\textwidth}
\centering
\includegraphics[width=.99\textwidth]{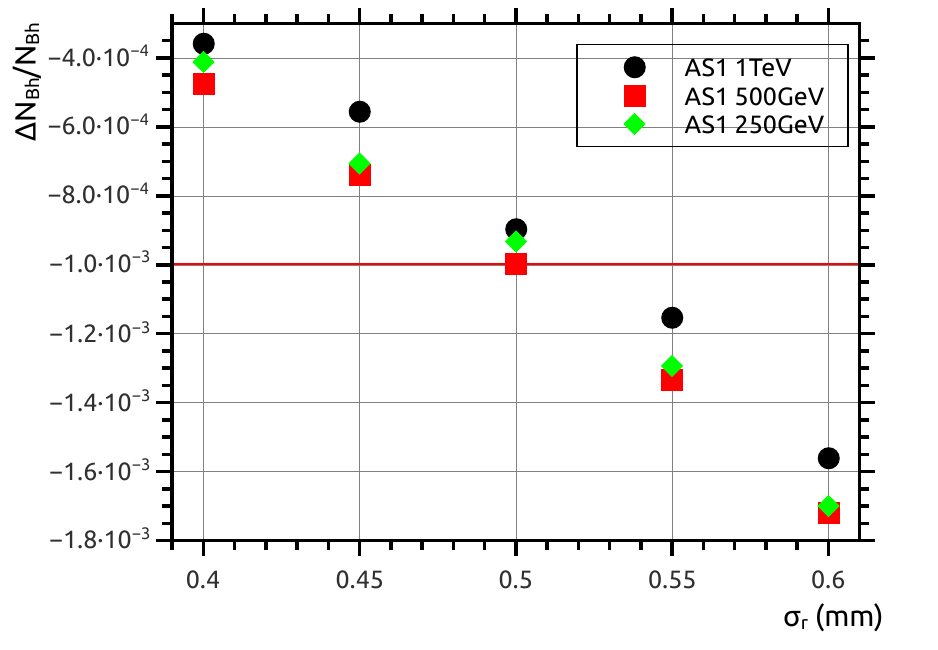}   
\caption{}
\end{subfigure}
\caption{\label{fig5} (a) Impact of the uncertainty $\sigma_{r}$ of measured shower radial position with respect to the true Bhabha position, at 91.2 GeV; (b) The same at 250 GeV, 500 GeV and 1 TeV center-of-mass energies for AS1 event selection (Section \ref{sec:sec2.2}).}
\end{figure}

\begin{figure}[!h]
\centering 
\begin{subfigure}[b]{0.47\textwidth}
\centering
\includegraphics[width=.99\textwidth]{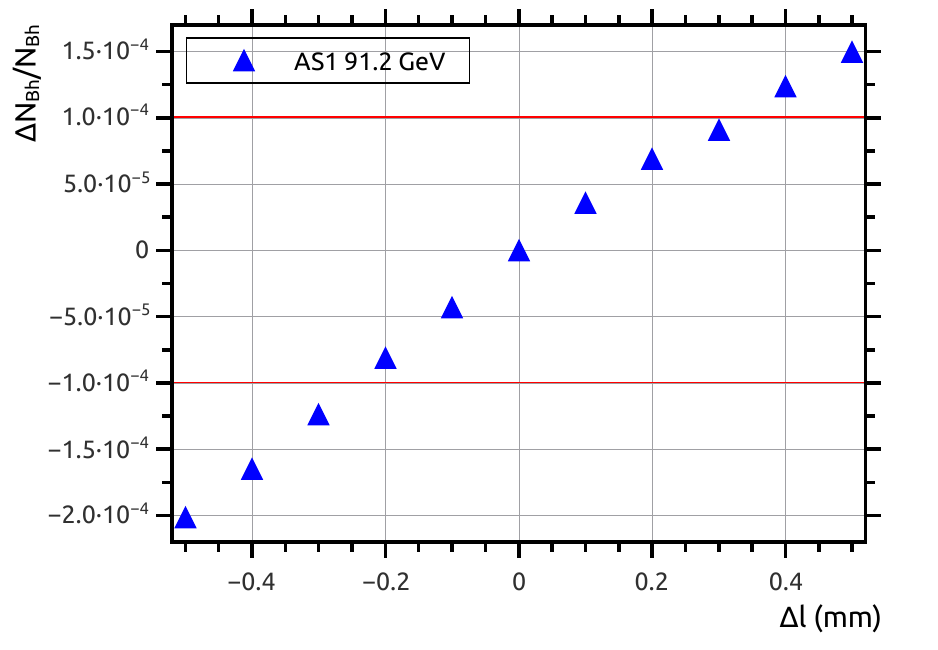}
\caption{}
\end{subfigure}
     \hfill
     \begin{subfigure}[b]{0.47\textwidth}
\centering
\includegraphics[width=.99\textwidth]{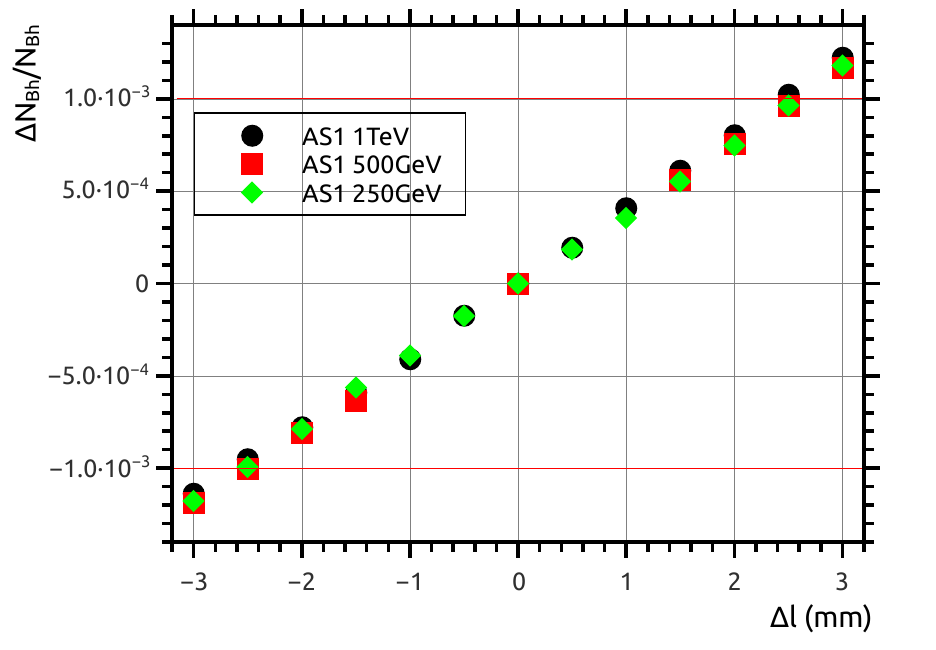}   
\caption{}
\end{subfigure}
\caption{\label{fig6} (a) Impact of uncertainty of the longitudinal distance $\Delta l$  between left and right halves of LumiCal, at 91.2 GeV;  (b) The same at 250 GeV, 500 GeV and 1 TeV center-of-mass energies for AS1 event selection (Section \ref{sec:sec2.2}). }
\end{figure}

\begin{figure}[!h]
\centering 
\begin{subfigure}[b]{0.455\textwidth}
\centering
\includegraphics[width=.99\textwidth]{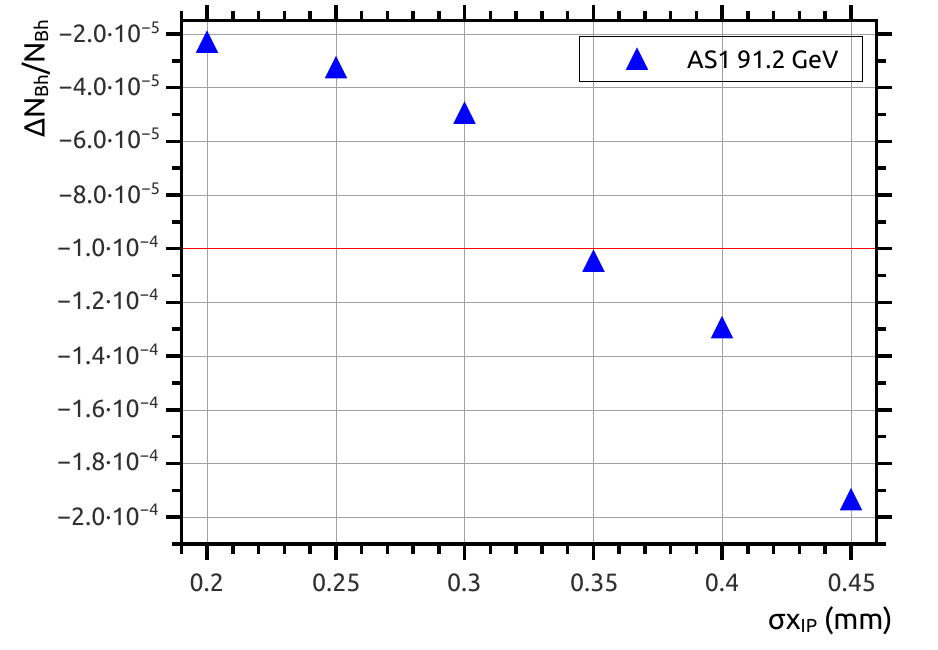}
\caption{}
\end{subfigure}
     \hfill
     \begin{subfigure}[b]{0.48\textwidth}
\centering
\includegraphics[width=.99\textwidth]{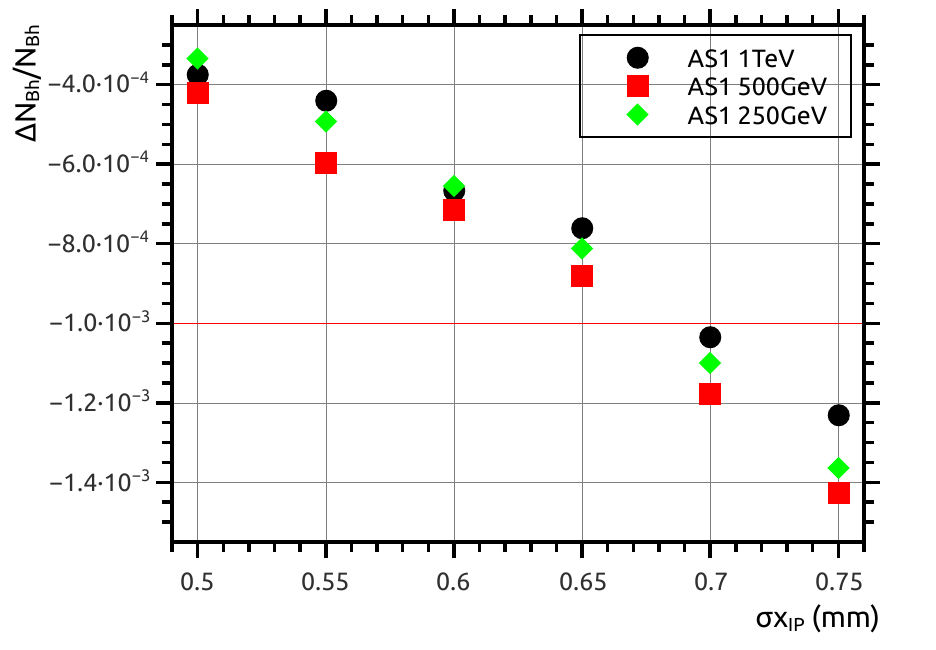}   
\caption{}
\end{subfigure}
\caption{\label{fig7} (a) Impact of radial fluctuations $\sigma_{x_{IP}}$ of LumiCal with respect to the IP, at 91.2 GeV; (b) The same at 250 GeV, 500 GeV and 1 TeV center-of-mass energies for AS1 event selection (Section \ref{sec:sec2.2}). }
\end{figure}

\begin{figure}[!h]
\centering 
\begin{subfigure}[b]{0.47\textwidth}
\centering
\includegraphics[width=.99\textwidth]{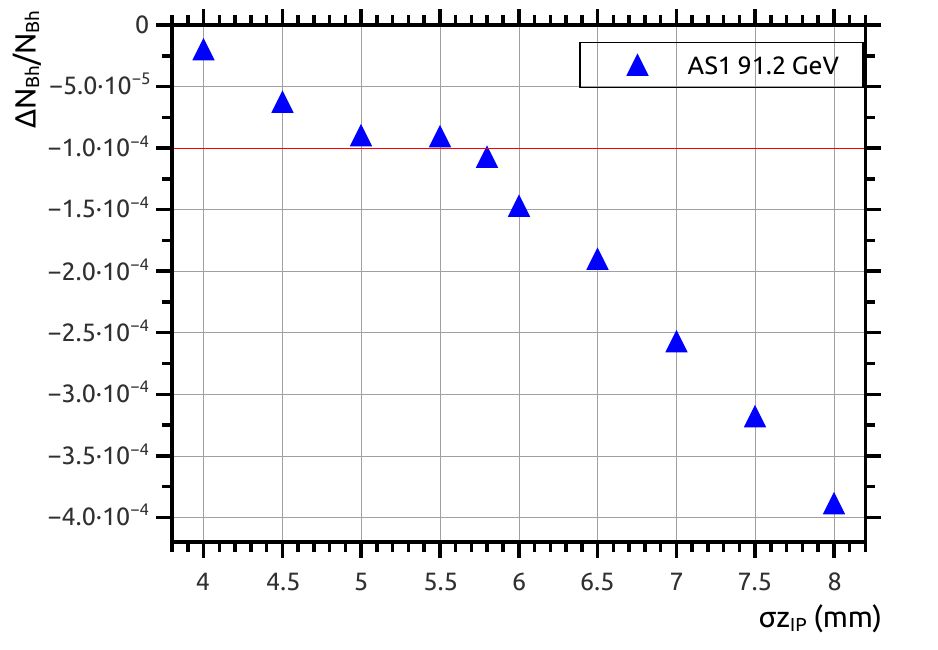}
\caption{}
\end{subfigure}
     \hfill
     \begin{subfigure}[b]{0.47\textwidth}
\centering
\includegraphics[width=.99\textwidth]{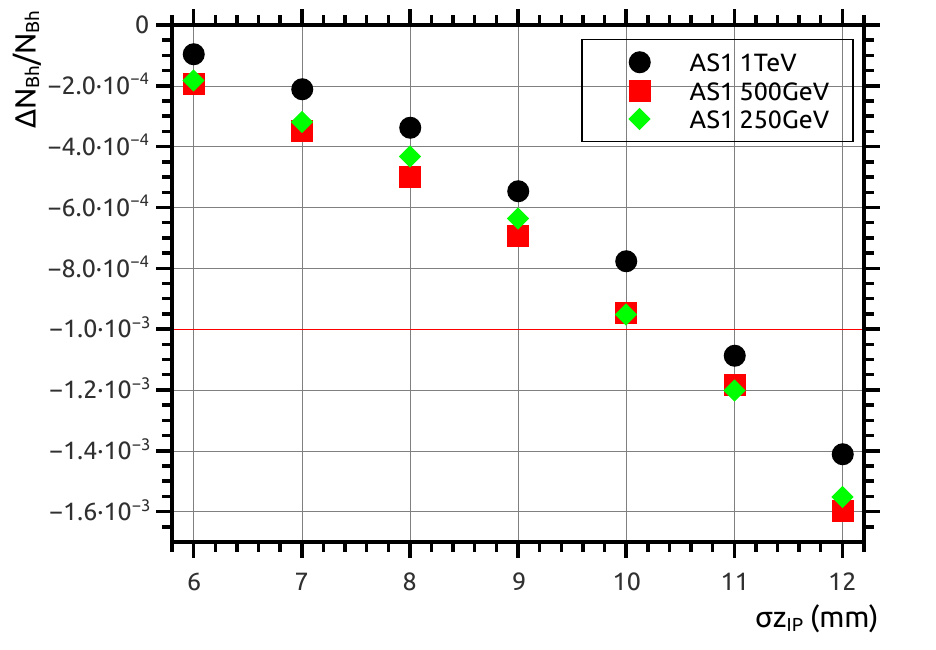}   
\caption{}
\end{subfigure}
\caption{\label{fig8} (a) Impact of axial fluctuations $\sigma_{z_{IP}}$ of LumiCal with respect to the IP, at 91.2 GeV; (b) The same at 250 GeV, 500 GeV and 1 TeV center-of-mass  energies for AS1 event selection (Section \ref{sec:sec2.2}). }
\end{figure}

\begin{figure}[!h]
\centering 
\begin{subfigure}[b]{0.47\textwidth}
\centering
\includegraphics[width=.99\textwidth]{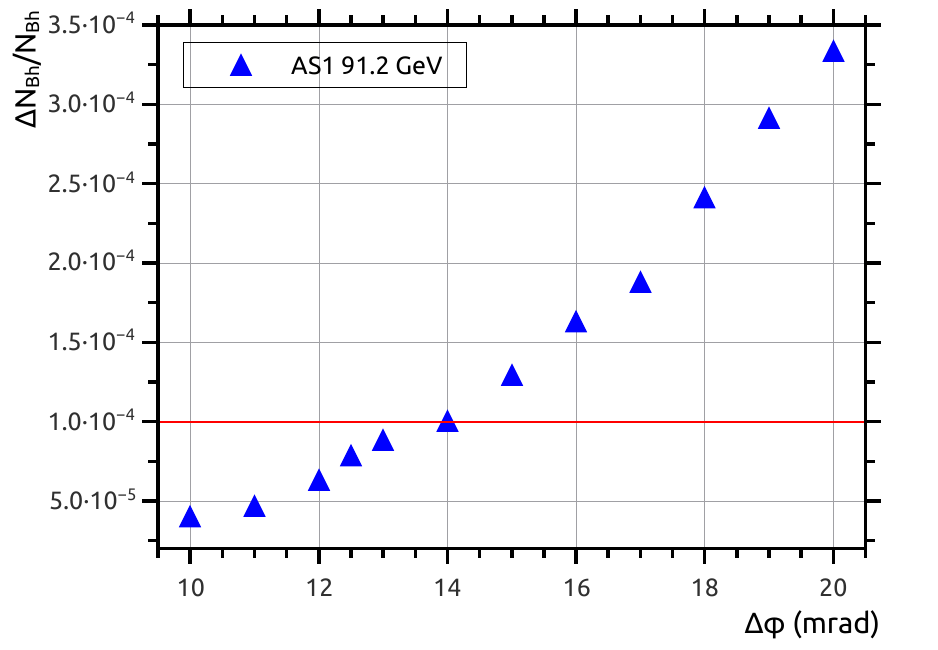}
\caption{}
\end{subfigure}
     \hfill
     \begin{subfigure}[b]{0.47\textwidth}
\centering
\includegraphics[width=.99\textwidth]{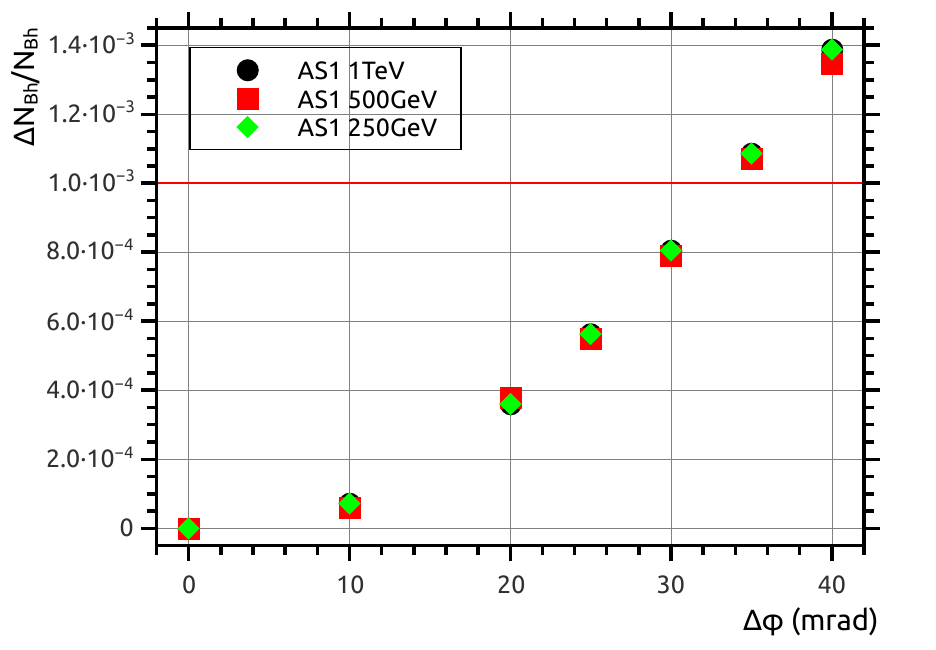}   
\caption{}
\end{subfigure}
\caption{\label{fig9} (a) Impact of rotation of LumiCal arms around the y-axis for a certain angle $\mathrm{\Delta \varphi}$ (tilt), at 91.2 GeV; (b) The same at 250 GeV, 500 GeV and 1 TeV center-of-mass energies for AS1 event selection (Section \ref{sec:sec2.2}). }
\end{figure}

\begin{figure}[!h]
\centering 
\begin{subfigure}[b]{0.47\textwidth}
\centering
\includegraphics[width=.99\textwidth]{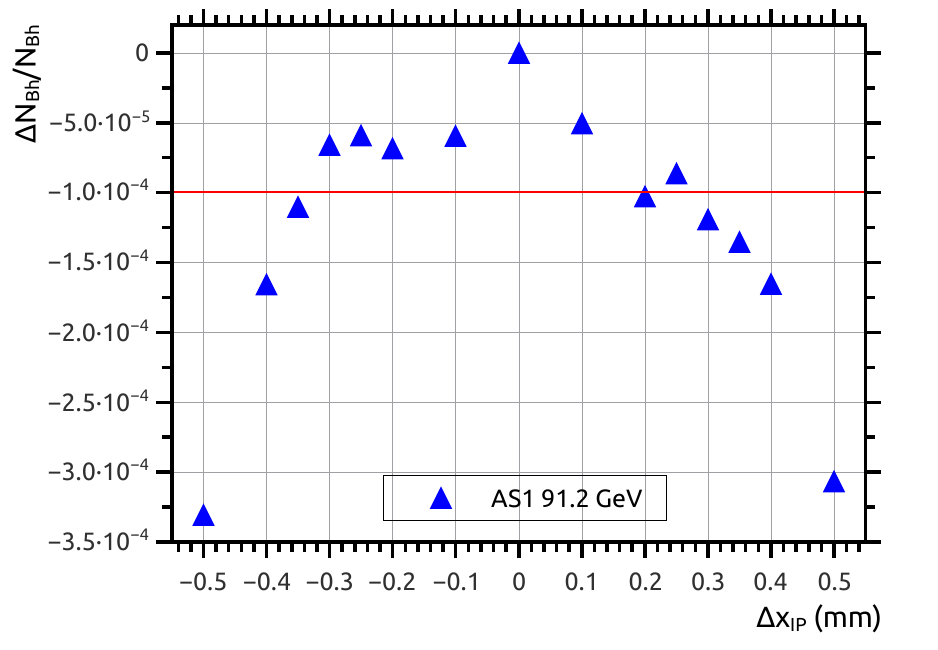}
\caption{}
\end{subfigure}
     \hfill
     \begin{subfigure}[b]{0.47\textwidth}
\centering
\includegraphics[width=.99\textwidth]{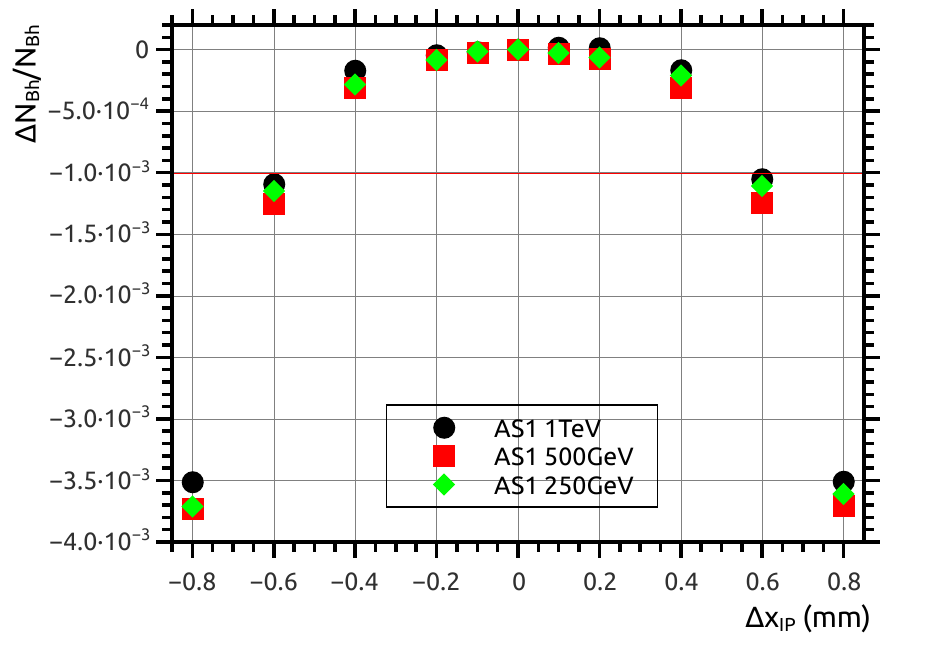}   
\caption{}
\end{subfigure}
\caption{\label{fig10} (a) Impact of radial displacements $\Delta_{x_{IP}}$ of the interaction point with respect to LumiCal, at 91.2 GeV;  (b) The same at 250 GeV, 500 GeV and 1 TeV center-of-mass energies for AS1 event selection (Section \ref{sec:sec2.2}). }
\end{figure}

\begin{figure}[!h]
\centering 
\begin{subfigure}[b]{0.47\textwidth}
\centering
\includegraphics[width=.99\textwidth]{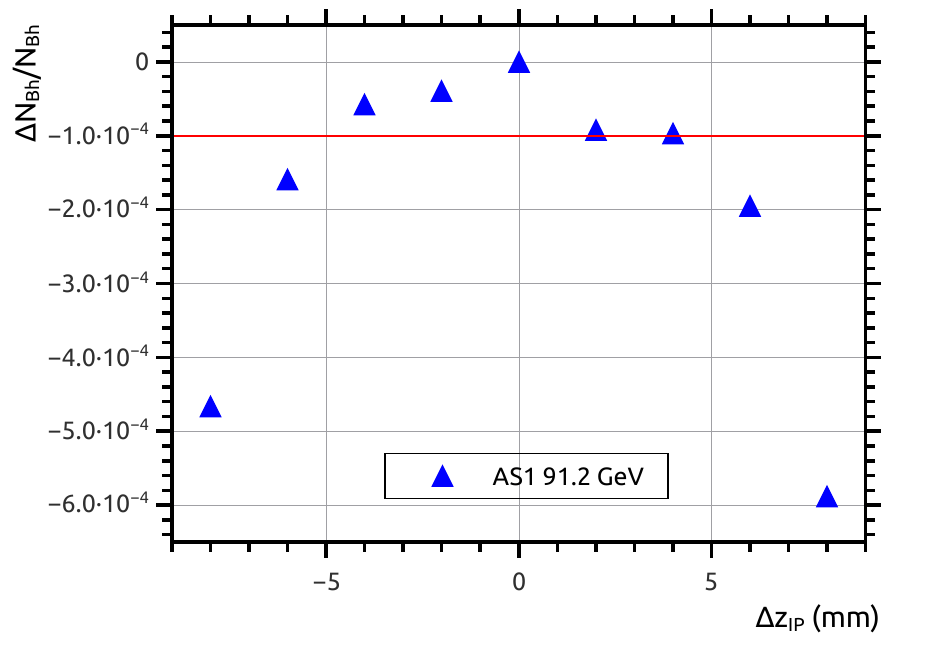}
\caption{}
\end{subfigure}
     \hfill
     \begin{subfigure}[b]{0.47\textwidth}
\centering
\includegraphics[width=.99\textwidth]{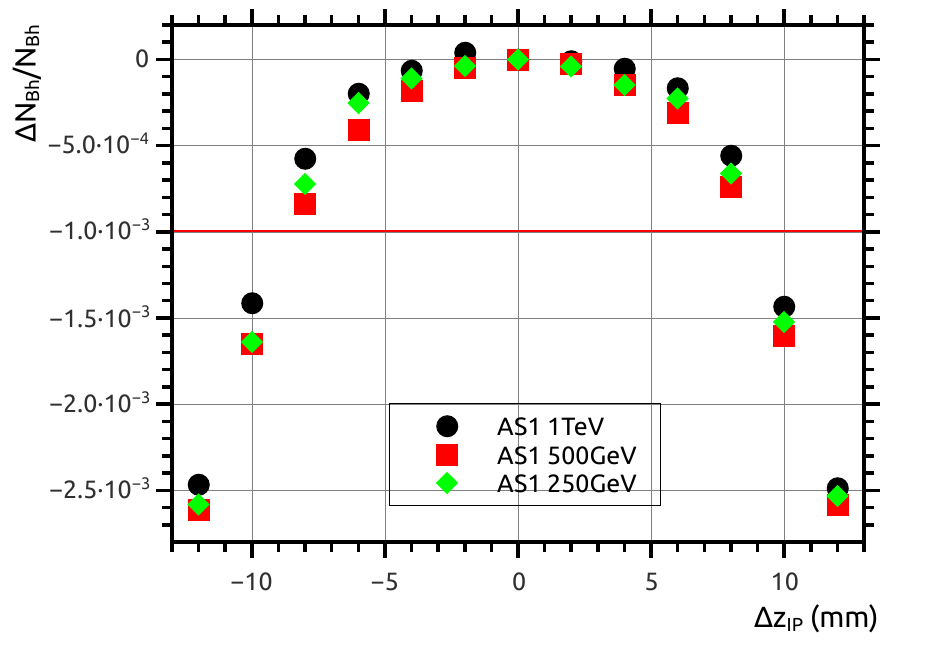}   
\caption{}
\end{subfigure}
\caption{\label{fig11} (a) Impact of axial displacements $\Delta z_{IP}$ of the interaction point with respect to LumiCal, at 91.2 GeV; (b) The same at 250 GeV, 500 GeV and 1 TeV center-of-mass energies for AS1 event selection (Section \ref{sec:sec2.2}). }
\end{figure}

\begin{figure}[!h]
\centering 
\begin{subfigure}[b]{0.47\textwidth}
\centering
\includegraphics[width=.99\textwidth]{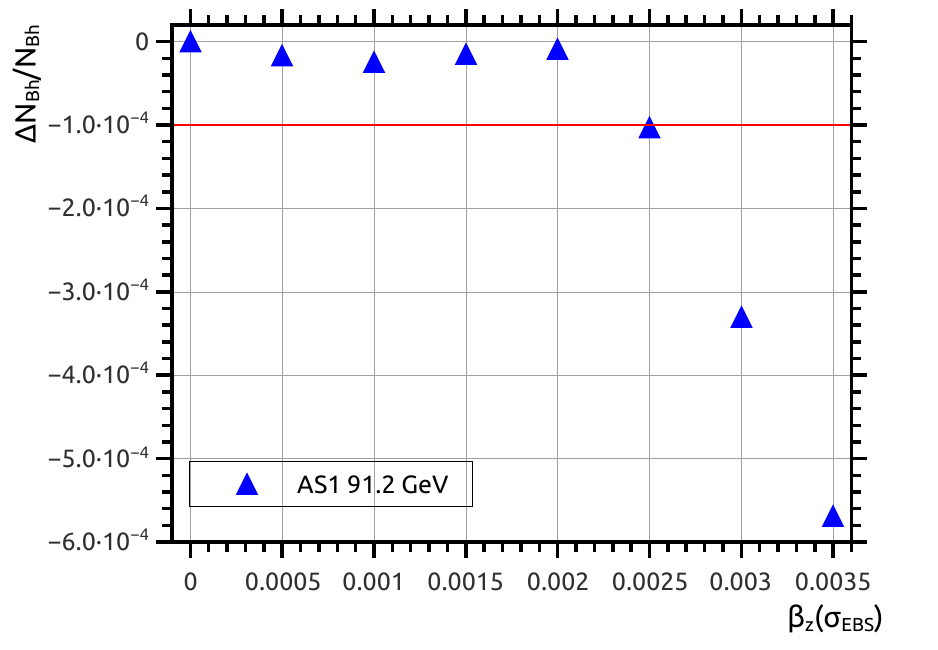}
\caption{}
\end{subfigure}
     \hfill
     \begin{subfigure}[b]{0.47\textwidth}
\centering
\includegraphics[width=.99\textwidth]{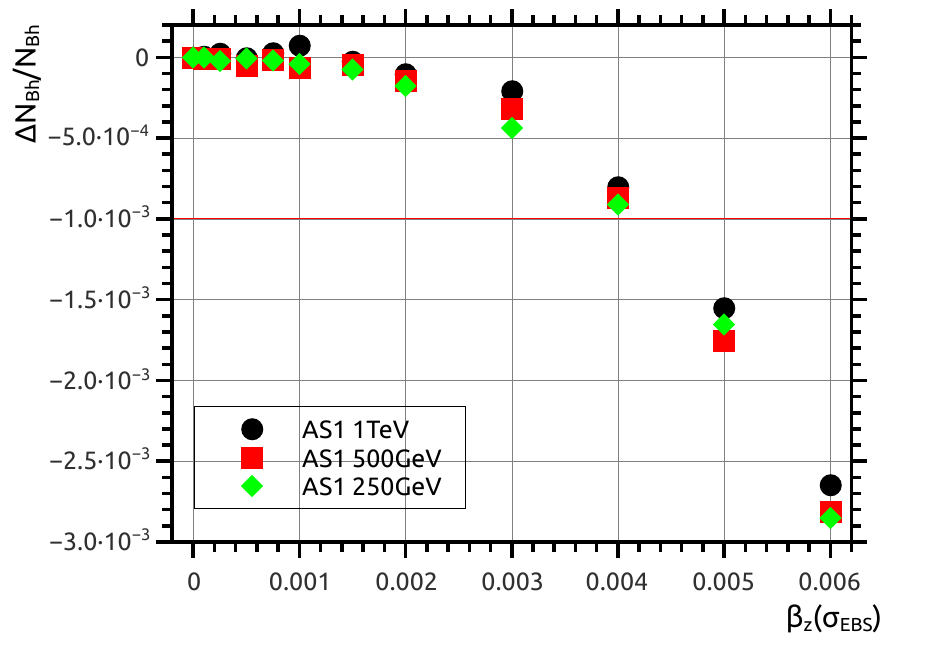}   
\caption{}
\end{subfigure}
\caption{\label{fig12} (a) Impact of the half-width $\sigma_{E_{BS}}$ of the Gaussian beam energy spread at 91.2 GeV as a source of longitudinal boost $\beta_z$ of the final state; (b) The same at 250 GeV, 500 GeV and 1 TeV center-of-mass energies for AS1 event selection (Section \ref{sec:sec2.2}).  }
\end{figure}

\begin{figure}[!h]
\centering 
\begin{subfigure}[b]{0.47\textwidth}
\centering
\includegraphics[width=.99\textwidth]{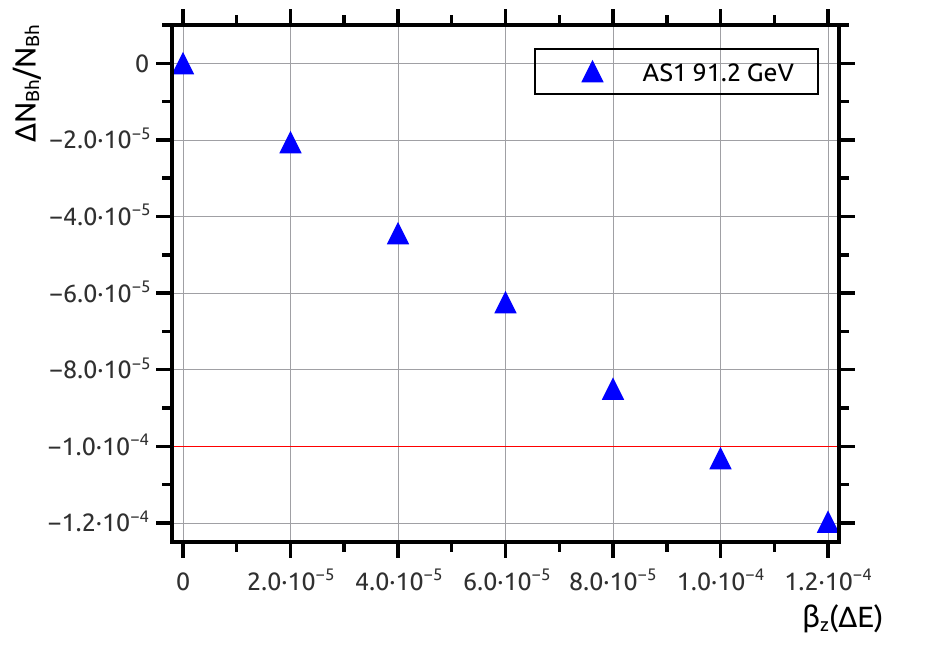}
\caption{}
\end{subfigure}
     \hfill
     \begin{subfigure}[b]{0.47\textwidth}
\centering
\includegraphics[width=.99\textwidth]{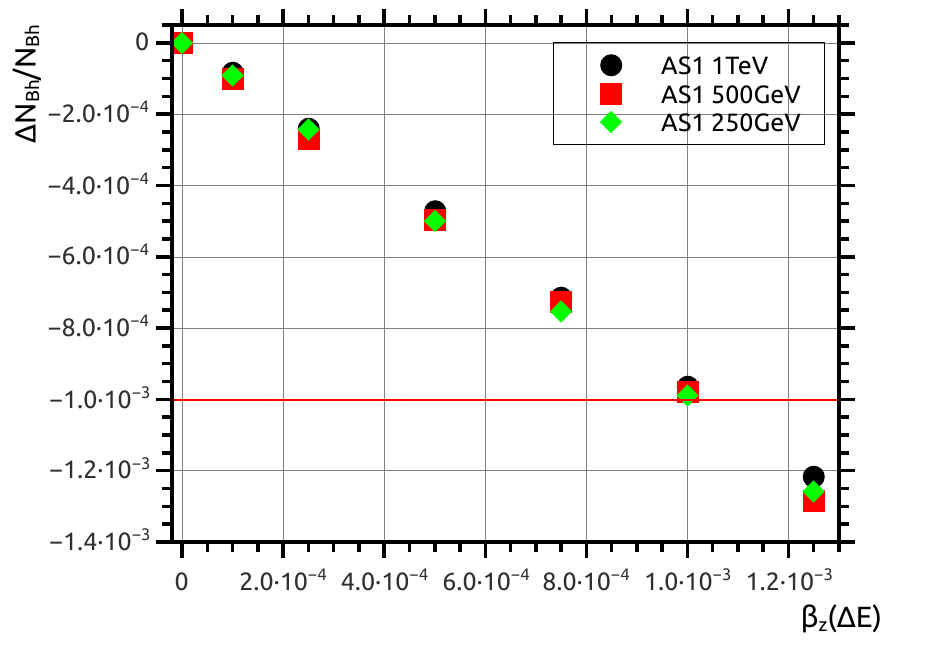}   
\caption{}
\end{subfigure}
\caption{\label{fig13}(a) Impact of the difference in energy $\Delta E$ between the colliding beams at 91.2 GeV, producing the longitudinal boost $\beta_z$; (b) The same at 250 GeV, 500 GeV and 1 TeV center-of-mass energies for AS1 event selection (Section \ref{sec:sec2.2}). }
\end{figure}

Similarly to the the situation at 91.2 GeV center-of-mass energy, Figures \ref{fig3} (b) to \ref{fig13} (b) illustrate the dependences of the relative change of Bhabha count ($\Delta N/N$) on the metrology effects listed as 1 to 11 respectively. All three center-of-mass energies share the same precision requirement on the relative systematic uncertainty of the integrated luminosity of $10^{-3}$. Dependences and limits on mechanical and beam parameters are similar at the considered center-of-mass energies, so they are collectively given at the same figure per each effect. The size of the uncertainties are given in regions of interest for the targeted precision of $\delta \mathcal{L}$ of $10^{-3}$ for each individual effect. Also, one should assume that the SABS as a calibration process does not receive any new physics contribution at high center-of-mass energies. 

As can been seen from Figures \ref{fig3} (b) to \ref{fig13} (b), precision requirements for metrology are typically relaxed a few times and up to an order of magnitude with respect to the ones at 91.2 GeV center-of-mass energy. The tilt of LumiCal is tolerable below $\sim$ 35 mrad (Figure \ref{fig9} (b)) which should be a feasible precision of the angular position control with the laser interferometry based systems. Yet, ILC, as well as the other future Higgs factories, requires a detailed study of a customized solution for position monitoring of subdetectors in the very forward region. An interesting question to be addressed in addition is what will be the impact of the push-pull option, if realized \cite{16}, on the precision estimates presented in this paper. 

The beam related requirements allow IP axial displacements up to 9 mm corresponding to a $\sim 30 \: \mathrm{ps}$ time-shift in beam synchronization, while radial displacements of the IP can be tolerated up to $\sim 600 \: \mathrm{\mu m}$ which is a very large tolerance with respect to the vertical size of the ILC nano-beams of 5.9 nm (at 250 GeV) \cite{16}. The beam energy spread smaller than $\mathrm{500 \: MeV}$ (at 250 GeV ILC) will not affect the integrated luminosity measurement in a relevant way, and this margin is two times larger than the beam spread in the current ILC design of $\sim$ 0.19\% \cite{16}. The same holds for higher center-of-mass energies. If an eventual difference of one beam energy with respect to the other would be present, it should not be larger than 1 permille of the beam-energy (Figure \ref{fig13} (b)) which also seems to be realistic.

\begin{table}[!h]
\caption{Maximal absolute precision of luminometer mechanical parameters and beam parameters, each contributing as $10^{-4}$($10^{-3}$) to the relative uncertainty of $\mathcal L$ at the $Z^{0}$ pole (higher energies). Values are approximated from Figures \ref{fig3} to \ref{fig13}.}
\label{table1}
\centering
\begin{tabular}{|c|c|c|c|c|}
\hline
\textbf{parameter} & \textbf{91.2 GeV} & \textbf{250 GeV} & \textbf{500 GeV} & \textbf{1 TeV}\\ 
\hline
$\Delta r_{in}$ ($\mathrm{\mu}$m) & 20 &  200 & 200 & 200\\
\hline
$\Delta r_{out}$ ($\mathrm{\mu}$m) & 60 &  600 & 600 & 550\\
\hline
$\sigma_{r}$ (mm) & 0.3 & 0.5 & 0.5 & 0.5\\
\hline
$\Delta l$ (mm) & 0.2 & 2.5 & 2.5 & 2.5\\
\hline
$\sigma_{x_{IP}}$ (mm) & 0.3 & 0.6& 0.6& 0.6\\
\hline
$\sigma_{z_{IP}}$ (mm) & 5 & 10 & 10 & 10\\
\hline
$\Delta \varphi$ (mrad) & 14 & 35 &  35 &  35\\
\hline
$\Delta x_{IP}$ (mm) & 0.3 & 0.6 &  0.6 & 0.6\\
\hline
$\Delta z_{IP}$ (mm) & 4 & 8 & 8 & 8\\
\hline
$\Delta \tau$ (ps) & 13& 27& 27& 30\\
\hline
$\sigma_{E_{BS}}$ (MeV) & 110 & 500 & 1000 & 2000\\
\hline
$\Delta E$ (MeV) & 5 & 125 & 250 & 500\\
\hline
\end{tabular}
\end{table}

The approximate maximal tolerance of the metrology parameters under study, extracted from Figures \ref{fig3} to \ref{fig13} is listed in Table \ref{table1}.


\section{Conclusion}
\label{sec:sec4}
This paper presents the first review of metrology effects in integrated luminosity measurements at ILC. We consider ILC operating energies at 91.2 GeV, 250 GeV, 500 GeV and 1 TeV center-of-mass energy. Each individual effect is assumed to contribute to the relative uncertainty of the integrated luminosity at the level of $10^{-4}$ and $10^{-3}$ at 91.2 GeV and higher energies, respectively, so the overall relative systematic uncertainty from metrology is limited to $3.3 \cdot10^{-4}$ and $3.3 \cdot10^{-3}$ in the worst case scenario. Since the precision achievable with the existing technologies, and in particular those related to the positioning and the alignment of LumiCal will be much higher than established here, realistic precision of $\mathcal{L}$ from metrology at ILC should be between $1 \cdot10^{-4} (10^{-3})$ and $3 \cdot10^{-4} (10^{-3})$ at the Z-pole and higher energies respectively.

This paper quantifies the metrology requirements to achieve the targeted precision and provide the guideline margins for the monitoring systems under development at ILC. At 91.2 GeV a few challenges are identified, such as the uncertainty of the available center-of-mass energy and the uncertainty of LumiCal inner aperture that should be known within $ 20 \: \mathrm{\mu m}$. It is yet to be confirmed in a full detector simulation how the uncertainty of LumiCal's inner aperture translates to the definition of the fiducial volume. At studied center-of-mass energies of 250 GeV and above we do not identify critical aspects of metrology from the point of view of existing technologies.


\section*{Acknowledgment}
This research was funded by the Ministry of Education, Science and Technological Development of the Republic of Serbia and by the Science Fund of the Republic of Serbia through the Grant No. 7699827, IDEJE HIGHTONE-P.
The study is done within the framework of ILD Concept Group. The authors would like to thank to the ILD Publication and Speakers Bureau and professor Graham W. Wilson from the Department of Physics and Astronomy, University of Kansas, for very useful comments and suggestions regarding this paper.


%

\vspace{0.2cm}
\noindent


\let\doi\relax


\end{document}